\documentclass{raa}
\usepackage{amsmath}
\usepackage{latexsym}
\usepackage{amssymb}
\usepackage{color,graphicx}
\usepackage{anysize}
\usepackage[sort]{natbib}
\usepackage{psfrag}
\newcommand{\bea}{\begin{eqnarray}}
\newcommand{\eea}{\end{eqnarray}}
\begin{document}

\title{Bianchi Type V Universe with Bulk Viscous Matter and Time Varying Gravitational and Cosmological Constants}

   \volnopage{Vol.0 (201x) No.0, 000--000}      
   \setcounter{page}{1}          

\author{Prashant S. Baghel$^*$
      \inst{}
\and J. P. Singh$^\dag$
      \inst{}
}

   \institute{Department of Mathematical Sciences, A. P. S. University, Rewa - 486003, India; $^*${\it ps.baghel@yahoo.com},~$^\dag${\it jpsinghmp@lycos.com}
}

\abstract{ Spatially homogeneous and anisotropic Bianchi type V space-time with bulk viscous fluid source and time varying gravitational constant $G$ and cosmological term $\Lambda$  are considered. Coefficient of bulk viscosity $\zeta$ is assumed as a simple linear function of Hubble parameter $H$ (i.e. $\zeta=\zeta_0+\zeta_1 H$, where $\zeta_0$ and $\zeta_1$ are constants). The Einstein field equations are solved explicitly by using a law of variation for the Hubble parameter, which yields a constant value of deceleration parameter. Physical and kinematical parameters of the models are discussed. The models are found to be compatible with the results of astronomical observations.
}
\authorrunning{P. S. Baghel \& J. P. Singh}            
\titlerunning{Bianchi type V universe with bulk viscous matter and time varying constants...}  
\maketitle
\section{Introduction}
In Einstein theory of Gravity Newtonian gravitational constant $G$ and Cosmological term $\Lambda$ are considered as fundamental constants. The Newtonian constant of gravitation $G$ plays the role of coupling constant between geometry of space and matter in Einstein field equations. In an evolving universe it is natural to look at this constant as a function of time. Ever since~\citet{rf301} first considered the possibility of variable $G$, there has been numerous modifications of general relativity to allow a variable $G$~\citep{rf302}. Nevertheless, these theories have not gained wide acceptance. However, recently a modification~\citep{rf303,rf304,rf305,rf306}  has been proposed in Einstein field equations treating $G$ and $\Lambda$ as coupling variables within the framework of general relativity.~\citet{rf307} have shown that the $G$ varying cosmology is consistent with whatever cosmological observations are available.~\citet{rf304,rf308,rf309} have discussed the possibility of an increasing $G$ .

Cosmological models with a cosmological constant are currently serious candidates to describe the expansion history of the universe. The huge difference between the small cosmological constant inferred from observations and vacuum energy density resulting from quantum field theories has been for along time, a difficult and fascinating problem~\citep{rf310,rf311,rf312,rf313} for cosmologists and field theory researchers. A wide range of observations~\citep{rf314,rf315,rf316,rf317} suggest that universe possesses a non-zero cosmological constant. At the same time, theories are increasingly exploring the possibility that this parameter is a dynamical one whose effective value in the early universe may have been quite different from the one we measure today. One possible explanation for a small $\Lambda$  term is to assume that it is dynamically evolving and not constant, that is, as the universe evolves from an earlier hotter and denser epoch, the effective cosmological term also evolves and decreases to its present value~\citep{rf318,rf319,rf320,rf321}.~\citet{rf322} obtained the time-dependent $G$ and $\Lambda$ solutions. Cosmological models with variable $G$ and $\Lambda$ term have been studied by a number of authors for homogeneous isotropic~\citep{rf309,rf323,rf324,rf325,rf326} and anisotropic~\citep{rf327,rf328,rf329,rf330,rfex1,rfex2,rfex3} space-times.

The investigations of relativistic cosmological models usually has the cosmic fluid as perfect fluid. However, these models do not incorporate dissipative mechanisms responsible for smoothing out initial anisotropies. It is believed that during neutrinos decoupling, the matter behaved like a viscous fluid~\citep{rf331} in the early stages of evolution.~\citet{rf332} studied Bianchi V viscous fluid cosmological models for barotropic  fluid distribution.~\citet{rf333,rf334,rf335} have investigated the role of viscosity in avoiding the initial big bang singularity.~\citet{rf336} investigated the effect of bulk viscosity on the evolution of the universe at large. They have shown that bulk viscosity leads to the inflationary like solutions.~\citet{rf337} discussed Bianchi type I universe with viscous fluid.~\citet{rfex4} investigated viscous fluid cosmological models in LRS Bianchi type V universe with varying $\Lambda$. Bulk viscous cosmological models with time-dependent $G$ and $\Lambda$ term have been studied by~\citet{rf329,rf338,rf339}, where bulk viscosity is taken as a power function of energy density. Recently~\citet{rf340} investigated Bianchi type V cosmological models with bulk viscosity, where the coefficient of bulk viscosity is assumed to be a power function of energy density $\rho$ or volume expansion $\theta$.

Among the physical quantities of interest in cosmology the deceleration parameter $q$ is currently a serious candidate to describe the dynamics of the universe. The prediction of standard cosmology, that the universe at present is decelerating is contradictory to the recent observational evidences of the high redshift of type Ia supernovae~\citep{rf341,rf342,rf343}. Observations reveal that instead of slowing down, the expanding universe is speeding up. Models with constant deceleration parameter have received considerable attention recently. The law of variation for Hubble parameter was initially proposed by~\citet{rf344} for FRW models that yields a constant value of deceleration parameter. Cosmological models with such a law of variation for Hubble parameter  have been studied by number of authors~\citep{rf345,rf346,rf347}. Recently~\citet{rf348,rf348a} proposed a similar law for the Hubble parameter and generated the solution for Bianchi type V space-time in general relativity. According to the proposed law the relation between the Hubble parameter $H$ and average scale factor $R$ is given by
\be\label{eq330}
H=\beta R^{-m},
\ee
where $\beta>0$  and $m\geq0$ are constants. For this variation law the deceleration parameter $q$ comes out to be constant, i.e.
\be\label{eq331}
q=m-1.
\ee
For $m>1$, the model represents a decelerating universe and $m<1$ corresponds to the accelerating phase of the universe. When $m=1$, we obtain $H\sim\frac{1}{t}$ and $q=0$. Therefore galaxies move with constant speed and model represents anisotropic Milne universe~\citep{rf349} for $m=1$. For $m=0$, we get $H=\beta$ and $q=-1$. Thus the observed Hubble parameter is true constant equal to its present value $H_0$ and model represents accelerating phase of the universe.

The relevance of the study of Bianchi type V cosmological models has already been discussed in our earlier papers~\citep{rf348,rf348a}, where we studied the viscous fluid models in some details. As a natural sequel to that study here we incorporate time varying $G$ and $\Lambda$ term in the bulk viscous Bianchi type V models and solve the coupled field equations exactly. In this paper we have taken the coefficient of bulk viscosity as simple linear function of Hubble parameter $H$~\citep{rf350}.

\section{Metric and Field Equations}
We consider Bianchi type V space-time in orthogonal form represented by the line-element
\be\label{eq301}
ds^2=-dt^2+A^2(t)dx^2+e^{2\alpha x}\left\{B^2(t)dy^2+C^2(t)dz^2\right\},
\ee
where $\alpha$ is a constant. We assume that the cosmic matter is represented by the energy-momentum tensor of an imperfect bulk viscous fluid
\be\label{eq302}
T_{ij}=(\rho+\bar{p})v_{i}v_{j}+\bar{p}g_{ij},
\ee
where $\bar{p}$ is the effective pressure given by
\be\label{eq303}
\bar{p}=p-\zeta {v^i}_{;i},
\ee
satisfying linear equation of state
\be\label{eq304}
p=\omega\rho.
\ee
Here $p$ is the equilibrium pressure, $\rho$ is the energy density of matter,  $\zeta$ is the coefficient of bulk viscosity and  $v^{i}$ is the flow vector of the fluid satisfying $v_iv^i=-1$. The semicolon stands for covariant differentiation. On thermodynamical grounds bulk viscous coefficient $\zeta$ is positive, assuring that the viscosity pushes the dissipative pressure $\bar{p}$ towards negative values. But correction to the thermodynamical pressure $p$ due to bulk viscous pressure is very small. Therefore, the dynamics of cosmic evolution does not change fundamentally by the inclusion of viscous term in the energy-momentum tensor.

The Einstein fields equations with time-dependent cosmological term $\Lambda$ and gravitational constant $G$ are
\be\label{eq305}
R_{ij}-\frac{1}{2}R_{k}^{k} g_{ij}=-8\pi G(t) T_{ij}+\Lambda(t) g_{ij}.
\ee
For the metric (\ref{eq301}) and matter distribution (\ref{eq302}) in comoving system of coordinates ($v_i=-\delta_{i}^{4}$), the field equations (\ref{eq305}) yield
\be\label{eq306}
8\pi G\bar{p}-\Lambda=\frac{\alpha^2}{A^2}-\frac{\ddot{B}}B-\frac{\ddot{C}}C-\frac{\dot{B}\dot{C}}{BC},
\ee
\be\label{eq307}
8\pi G\bar{p}-\Lambda=\frac{\alpha^2}{A^2}-\frac{\ddot{C}}C-\frac{\ddot{A}}A-\frac{\dot{C}\dot{A}}{CA},
\ee
\be\label{eq308}
8\pi G\bar{p}-\Lambda=\frac{\alpha^2}{A^2}-\frac{\ddot{A}}A-\frac{\ddot{B}}B-\frac{\dot{A}\dot{B}}{AB},
\ee
\be\label{eq309}
8\pi G\rho+\Lambda=-\frac{3\alpha^2}{A^2}+\frac{\dot{A}\dot{B}}{AB}+\frac{\dot{B}\dot{C}}{BC}+\frac{\dot{A}\dot{C}}{AC},
\ee
\be\label{eq310}
0=\frac{2\dot{A}}{A}-\frac{\dot{B}}{B}-\frac{\dot{C}}{C},
\ee
where an overhead dot (.) denotes ordinary differentiation with respect to cosmic time $t$. Due to contracted Bianchi identity, divergence of Einstein tensor $G_{ij}=R_{ij}-\frac{1}{2}R_{k}^{k} g_{ij}$ is zero and we get
\be\label{eq311}
8\pi G\left\{\dot{\rho}+(\rho+\bar{p})\left(\frac{\dot{A}}{A}+\frac{\dot{B}}{B}+\frac{\dot{C}}{C}\right)\right\}+8\pi\dot{G}+\dot{\Lambda}=0.
\ee
The usual conservation equation ${T_{i}^{j}}_{;j}=0$ splits above equation into
\be\label{eq312}
\dot{\rho}+(\rho+\bar{p})\left(\frac{\dot{A}}{A}+\frac{\dot{B}}{B}+\frac{\dot{C}}{C}\right)=0
\ee
and
\be\label{eq313}
8\pi\rho\dot{G}+\dot{\Lambda}=0.
\ee
From (\ref{eq313}), one concludes that, when $\Lambda$ is constant or zero, $G$ turns out to be constant for non-zero energy density.

We define the average scale factor $R$ for Bianchi V space-time as $R^3=ABC$. From equations (\ref{eq306})-(\ref{eq308}) and (\ref{eq310}), we obtain
\be\label{eq314}
\frac{\dot{A}}{A}=\frac{\dot{R}}{R},
\ee
\be\label{eq315}
\frac{\dot{B}}{B}=\frac{\dot{R}}{R}-\frac{k}{R^3},
\ee
\be\label{eq316}
\frac{\dot{C}}{C}=\frac{\dot{R}}{R}+\frac{k}{R^3},
\ee
$k$ being constant of integration. On integration, equations (\ref{eq314})-(\ref{eq316}) give
\be\label{eq317}
A=m_1R,
\ee
\be\label{eq318}
B=m_2R\exp\left(-k\int\frac{dt}{R^3}\right),
\ee
\be\label{eq319}
C=m_3R\exp\left(k\int\frac{dt}{R^3}\right),
\ee
where $m_1$, $m_2$ and $m_3$ are constants of integration satisfying $m_1m_2m_3=1$.

In analogy with FRW universe, we define a generalized Hubble parameter $H$ and generalized deceleration parameter $q$ as
\be\label{eq320}
H=\frac{\dot{R}}{R}=\frac{1}{3}(H_1+H_2+H_3)
\ee
and
\be\label{eq321}
q=-1-\frac{\dot{H}}{H^2}
\ee
where $H_1=\frac{\dot{A}}{A}$, $H_2=\frac{\dot{B}}{B}$ and $H_3=\frac{\dot{C}}{C} $ are directional Hubble's factors along $x$, $y$ and $z$ directions respectively.

We introduce volume expansion $\theta$ and shear $\sigma$, as usual
$$
\theta=v^i_{;i}~\mbox{and}~\sigma^2=\frac{1}{2}\sigma_{ij}\sigma^{ij},
$$
$\sigma^{ij}$ being shear tensor. For the Bianchi V metric, expressions  for $\theta$ and $\sigma$ come out to be
\be\label{eq322}
\theta=\frac{3\dot{R}}{R}
\ee
and
\be\label{eq323}
\sigma=\frac{k}{R^3}.
\ee

We can express equations (\ref{eq306})-(\ref{eq309}) and (\ref{eq312}) in the terms of $R$, $H$, $q$ and $\sigma$ as
\be\label{eq326}
8\pi G\bar{p}-\Lambda=(2q-1)H^2-\sigma^2+\frac{\alpha^2}{R^2},
\ee
\be\label{eq327}
8\pi G\rho+\Lambda=3H^2-\sigma^2-\frac{3\alpha^2}{R^2},
\ee
\be\label{eq328}
\dot{\rho}+3(\rho+\bar{p})H=0.
\ee
It is to note that energy density of the universe is a positive quantity. It is believed that at the early stages of evolution, when the average scale factor $R$ was close to zero, the energy density of the universe was infinitely large. On the other hand, with expansion of the universe i.e. with increase of $R$ , the energy density decreases and an infinitely large $R$ corresponds to $\rho$ close to zero. In that case from (\ref{eq327}), we obtain $\frac{\rho_v}{\rho_c}\rightarrow 1$, where $\rho_v=\frac{\Lambda}{8\pi G}$ and $\rho_c=\frac{3H^2}{8\pi G}$. For $\Lambda\geq0$, $\rho<\rho_c$. Also from (\ref{eq327}), we observe that $0<\frac{\sigma^2}{\theta^2}<\frac{1}{3}$ and $0<\frac{8\pi G\rho}{\theta^2}<\frac{1}{3}$  for $\Lambda\geq0$. Thus a positive $\Lambda$ restricts the upper limit of anisotropy whereas a negative $\Lambda$ will increase the anisotropy.\\
From (\ref{eq326}) and (\ref{eq327}), we get
\be\label{eq329}
\frac{d\theta}{dt}=\Lambda+12\pi G\zeta\theta-4\pi G(\rho+3{p})-2\sigma^2-\frac{\theta^2}{3},
\ee
which is the Raychaudhuri equation for the given distribution. We observe that for negative $\Lambda$ and the absence of viscosity, the universe will always be in decelerating phase provided the strong energy conditions~\citep{rf351} hold whereas in the presence of viscosity, positive $\Lambda$   will slow down the rate of decrease of volume expansion. Also, $\dot{\sigma}=-\sigma\theta$ implying that $\sigma$ decreases in an evolving universe and for infinitely large value of $R$, $\sigma$ becomes negligible.

\section{Solution of the Field Equations}
Integrating (\ref{eq330}), we obtain
\be\label{eq332}
R=(m\beta t+t_1)^{\frac{1}{m}}~~~~\mbox{for $m\neq0$}
\ee
and
\be\label{eq333}
R=\exp\{\beta(t-t_0)\}~~~~\mbox{for $m=0$},
\ee
where $t_1$ and $t_0$ are constants of integration. From (\ref{eq317})-(\ref{eq319}) with the use of (\ref{eq332}), we obtain
\be\label{}
A=m_1(m\beta t+t_1)^{\frac{1}{m}},\notag
\ee
\be\label{}
B=m_2(m\beta t+t_1)^{\frac{1}{m}}\exp\left\{\frac{-k(m\beta t+t_1)^{\frac{m-3}{m}}}{\beta(m-3)}\right\},\notag
\ee
\be\label{}
C=m_3(m\beta t+t_1)^{\frac{1}{m}}\exp\left\{\frac{k(m\beta t+t_1)^{\frac{m-3}{m}}}{\beta(m-3)}\right\}.\notag
\ee
For this solution, the metric (\ref{eq301}) assumes the following form after suitable transformation of coordinates
\bea
ds^2&=&-dT^2+(m\beta T)^{\frac{2}{m}}dX^2+(m\beta T)^{\frac{2}{m}}\exp\left\{2\alpha X-\frac{2k(m\beta T)^{\frac{m-3}{m}}}{\beta(m-3)}\right\}dY^2\notag\\
\label{eq334}
& &+(m\beta T)^{\frac{2}{m}}\exp\left\{2\alpha X+\frac{2k(m\beta T)^{\frac{m-3}{m}}}{\beta(m-3)}\right\}dZ^2.
\eea
Equations (\ref{eq317})-(\ref{eq319}) together with (\ref{eq333}) give
\be\label{}
A=m_1\exp\{\beta(t-t_0)\},\notag
\ee
\be\label{}
B=m_2\exp\left\{\beta(t-t_0)+\frac{k}{3\beta}e^{-3\beta(t-t_0)}\right\},\notag
\ee
\be\label{}
C=m_3\exp\left\{\beta(t-t_0)-\frac{k}{3\beta}e^{-3\beta(t-t_0)}\right\}.\notag
\ee
The line-element (\ref{eq301}) for this solution can be written as
\bea
ds^2&=&-dT^2+e^{2\beta T}dX^2
+\exp\left(2\alpha X +2\beta T+\frac{2k}{3\beta}e^{-3\beta T}\right)dY^2\notag\\
\label{eq335}
& &+\exp\left(2\alpha X+2\beta T-\frac{2k}{3\beta}e^{-3\beta T}\right)dZ^2.
\eea

\section{Discussion}
To determine the coefficient of bulk viscosity, $\zeta$ is assumed to be a simple linear function of Hubble parameter $H$~\citep{rf350}, i.e.
\be\label{eq336}
\zeta=\zeta_0+\zeta_1H,
\ee
where $\zeta_0~(\geq0)$ and $\zeta_1$ are constants. For this choice, equation (\ref{eq328}) reduces to
\be\label{eq337}
\dot{\rho}+3(1+\omega)H\rho=9(\zeta_0+\zeta_1H)H^2.
\ee
We discuss the models for $m\neq0$ and $m=0$.
\subsection{Cosmology for $m\neq0$}
\begin{figure}
\def\PFGstyle{\tiny}
\centering
\begin{psfrags}
\psfrag{MatterEner}[bc][bc]{\PFGstyle Matter Energy Density $\rho $}%
\psfrag{mNotEqual0}[cc][cc]{\PFGstyle $m\neq0$}%
\psfrag{Omega0}[cl][cl]{\PFGstyle $\omega $=0}%
\psfrag{Omega13}[cl][cl]{\PFGstyle $\omega $=1/3}%
\psfrag{Omega1}[cl][cl]{\PFGstyle $\omega $=1}%
\psfrag{S0}[tc][tc]{\PFGstyle $0$}%
\psfrag{S12}[tc][tc]{\PFGstyle $10$}%
\psfrag{S21}[tc][tc]{\PFGstyle $2$}%
\psfrag{S41}[tc][tc]{\PFGstyle $4$}%
\psfrag{S61}[tc][tc]{\PFGstyle $6$}%
\psfrag{S81}[tc][tc]{\PFGstyle $8$}%
\psfrag{Timet}[tc][tc]{\PFGstyle Time $t$}%
\psfrag{W0}[cr][cr]{\PFGstyle $0$}%
\psfrag{W13}[cr][cr]{\PFGstyle $100$}%
\psfrag{W22}[cr][cr]{\PFGstyle $20$}%
\psfrag{W42}[cr][cr]{\PFGstyle $40$}%
\psfrag{W62}[cr][cr]{\PFGstyle $60$}%
\psfrag{W82}[cr][cr]{\PFGstyle $80$}%
\psfrag{x0}[tc][tc]{\PFGstyle $0$}%
\psfrag{x11}[tc][tc]{\PFGstyle $1$}%
\psfrag{x151}[tc][tc]{\PFGstyle $1.5$}%
\psfrag{x21}[tc][tc]{\PFGstyle $2$}%
\psfrag{x251}[tc][tc]{\PFGstyle $2.5$}%
\psfrag{x2}[tc][tc]{\PFGstyle $0.2$}%
\psfrag{x4}[tc][tc]{\PFGstyle $0.4$}%
\psfrag{x5}[tc][tc]{\PFGstyle $0.5$}%
\psfrag{x6}[tc][tc]{\PFGstyle $0.6$}%
\psfrag{x8}[tc][tc]{\PFGstyle $0.8$}%
\psfrag{xm11}[tc][tc]{\PFGstyle $-1$}%
\psfrag{xm5}[tc][tc]{\PFGstyle $-0.5$}%
\psfrag{y0}[cr][cr]{\PFGstyle $0$}%
\psfrag{y11}[cr][cr]{\PFGstyle $1$}%
\psfrag{y21}[cr][cr]{\PFGstyle $2$}%
\psfrag{y2}[cr][cr]{\PFGstyle $0.2$}%
\psfrag{y31}[cr][cr]{\PFGstyle $3$}%
\psfrag{y41}[cr][cr]{\PFGstyle $4$}%
\psfrag{y4}[cr][cr]{\PFGstyle $0.4$}%
\psfrag{y5}[cr][cr]{\PFGstyle $0.5$}%
\psfrag{y6}[cr][cr]{\PFGstyle $0.6$}%
\psfrag{ym11}[cr][cr]{\PFGstyle $-1$}%
\psfrag{ym2}[cr][cr]{\PFGstyle $-0.2$}%
\psfrag{ym4}[cr][cr]{\PFGstyle $-0.4$}%
\psfrag{ym5}[cr][cr]{\PFGstyle $-0.5$}%
\psfrag{ym6}[cr][cr]{\PFGstyle $-0.6$}%
\includegraphics[width=4in]{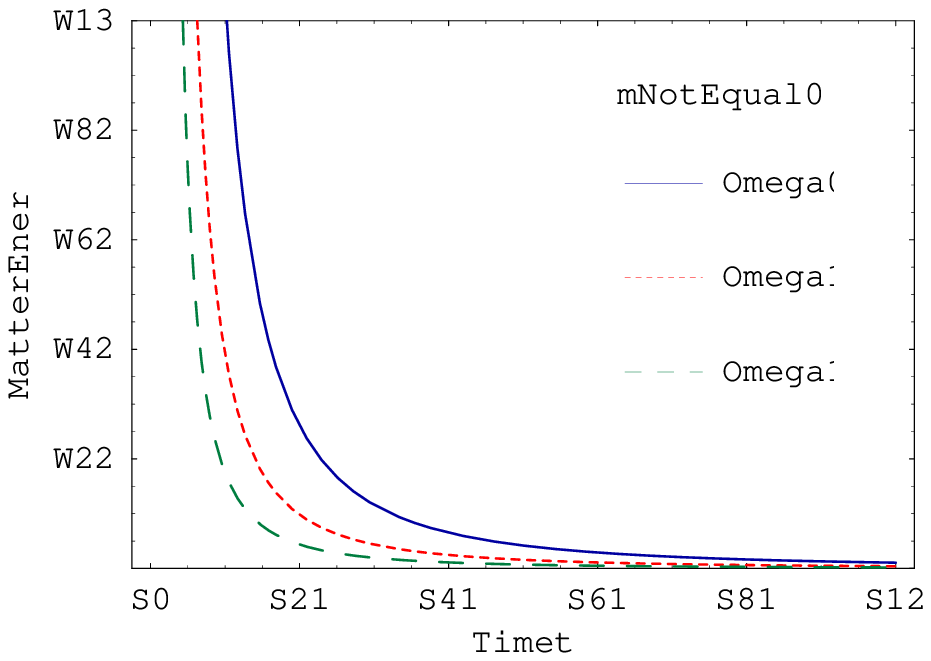}
\end{psfrags}
 \caption{Variation of matter energy density $\rho$ with cosmic time $t$ in case of $m\neq0$.}\protect\label{fig301}
\end{figure}
For the model (\ref{eq334}), average scale factor $R$ is given by
\be\label{eq338}
R=(m\beta T)^\frac{1}{m}.
\ee
Volume expansion $\theta$, Hubble parameter $H$ and shear scalar $\sigma$ for this model are
\be\label{eq339}
\theta=3H=\frac{3}{mT},
\ee
\be\label{eq340}
\sigma=k(m\beta T)^\frac{-3}{m}.
\ee
Using (\ref{eq339}) in (\ref{eq336}) and (\ref{eq337}), we obtain
\be\label{eq341}
\zeta=\zeta_0+\frac{\zeta_1}{mT},
\ee
\be\label{eq342}
\rho=\frac{9\zeta_0}{m(3+3\omega-m)T}+\frac{9\zeta_1}{(3+3\omega-2m)m^2T^2}+\frac{a}{T^{3(1+\omega)/m}},
\ee
where $a$ is an integration constant. Gravitational constant $G$ and cosmological term $\Lambda$ are obtained as
\be\label{eq343}
G=\frac{1}{4\pi}\left[\frac{mT^2\left\{\alpha^2(m\beta T)^\frac{4}{m}+k^2\right\}-(m\beta T)^\frac{6}{m}}{\left(\frac{3m\zeta_0}{m-3\omega-3}\right)T^\frac{m+6}{m}+\left(\frac{6\zeta_1}{2m-3\omega-3}\right)T^\frac{6}{m}-ma(1+\omega)T^\frac{2m+3(1-\omega)}{m}}\right],
\ee
\bea
\Lambda&=&18\left\{\frac{a}{9T^\frac{3(1+\omega)}{m}}+\frac{\zeta_0}{(3+3\omega-m)m T}+\frac{\zeta_1}{(3+3\omega-2m)m^2T^2}\right\}\notag\\
& &\cdot\left[\frac{(m\beta T)^\frac{6}{m}-mT^2\left\{\alpha^2(m\beta T)^\frac{4}{m}+k^2\right\}}{\left(\frac{3m\zeta_0}{m-3\omega-3}\right)T^\frac{m+6}{m}+\left(\frac{6\zeta_1}{2m-3\omega-3}\right)T^\frac{6}{m}-ma(1+\omega)T^\frac{2m+3(1-\omega)}{m}}\right]\notag\\
\label{eq344}
& &+\frac{3}{m^2T^2}-\frac{k^2}{(m\beta T)^\frac{6}{m}}-\frac{3\alpha^2}{(m\beta T)^\frac{2}{m}}.
\eea
We observe that the model is not tenable for $m=3$. The model has singularity at $T=0$. At $T=0$, $\rho$, $\Lambda$, $\zeta$, $\theta$, $\sigma$ all diverge whereas $G$  becomes a constant provided $m>3$. In the limit of large times, $\rho$, $\theta$, $\sigma$ become zero, $\zeta=\zeta_0$ and $G\rightarrow\infty$, $\Lambda\rightarrow$constant for $m>3$. Also for $T\rightarrow\infty$, $\sigma/\theta\rightarrow0$ when $m<3$. Therefore the model approaches isotropy asymptotically. For large values of $T$ the model becomes conformally flat~\citep{rf348}. The integral
\be\label{eq346}
\int_{T_0}^{T}\frac{dt}{R(t)}=\frac{1}{\beta(m-1)}\left[(m\beta T)^\frac{m-1}{m}\right]
\ee
is finite provided $m\neq1$ . Therefore particle horizon exists in the model. It is to note that for $m=1$, the model does not have horizon.
\subsection{Cosmology for $m=0$}
\begin{figure}
\def\PFGstyle{\tiny}
\centering
\begin{psfrags}
\psfrag{m0}[cc][cc]{\PFGstyle $m=0$}%
\psfrag{MatterEner}[bc][bc]{\PFGstyle Matter Energy Density $\rho $}%
\psfrag{Omega0}[cl][cl]{\PFGstyle $\omega $=0}%
\psfrag{Omega13}[cl][cl]{\PFGstyle $\omega $=1/3}%
\psfrag{Omega1}[cl][cl]{\PFGstyle $\omega $=1}%
\psfrag{S0}[tc][tc]{\PFGstyle $0$}%
\psfrag{S12}[tc][tc]{\PFGstyle $10$}%
\psfrag{S21}[tc][tc]{\PFGstyle $2$}%
\psfrag{S41}[tc][tc]{\PFGstyle $4$}%
\psfrag{S61}[tc][tc]{\PFGstyle $6$}%
\psfrag{S81}[tc][tc]{\PFGstyle $8$}%
\psfrag{Timet}[tc][tc]{\PFGstyle Time $t$}%
\psfrag{W0}[cr][cr]{\PFGstyle $0$}%
\psfrag{W11}[cr][cr]{\PFGstyle $1$}%
\psfrag{W2}[cr][cr]{\PFGstyle $0.2$}%
\psfrag{W4}[cr][cr]{\PFGstyle $0.4$}%
\psfrag{W6}[cr][cr]{\PFGstyle $0.6$}%
\psfrag{W8}[cr][cr]{\PFGstyle $0.8$}%
\psfrag{x0}[tc][tc]{\PFGstyle $0$}%
\psfrag{x11}[tc][tc]{\PFGstyle $1$}%
\psfrag{x151}[tc][tc]{\PFGstyle $1.5$}%
\psfrag{x21}[tc][tc]{\PFGstyle $2$}%
\psfrag{x251}[tc][tc]{\PFGstyle $2.5$}%
\psfrag{x2}[tc][tc]{\PFGstyle $0.2$}%
\psfrag{x4}[tc][tc]{\PFGstyle $0.4$}%
\psfrag{x5}[tc][tc]{\PFGstyle $0.5$}%
\psfrag{x6}[tc][tc]{\PFGstyle $0.6$}%
\psfrag{x8}[tc][tc]{\PFGstyle $0.8$}%
\psfrag{xm11}[tc][tc]{\PFGstyle $-1$}%
\psfrag{xm5}[tc][tc]{\PFGstyle $-0.5$}%
\psfrag{y0}[cr][cr]{\PFGstyle $0$}%
\psfrag{y11}[cr][cr]{\PFGstyle $1$}%
\psfrag{y21}[cr][cr]{\PFGstyle $2$}%
\psfrag{y2}[cr][cr]{\PFGstyle $0.2$}%
\psfrag{y31}[cr][cr]{\PFGstyle $3$}%
\psfrag{y41}[cr][cr]{\PFGstyle $4$}%
\psfrag{y4}[cr][cr]{\PFGstyle $0.4$}%
\psfrag{y5}[cr][cr]{\PFGstyle $0.5$}%
\psfrag{y6}[cr][cr]{\PFGstyle $0.6$}%
\psfrag{ym11}[cr][cr]{\PFGstyle $-1$}%
\psfrag{ym2}[cr][cr]{\PFGstyle $-0.2$}%
\psfrag{ym4}[cr][cr]{\PFGstyle $-0.4$}%
\psfrag{ym5}[cr][cr]{\PFGstyle $-0.5$}%
\psfrag{ym6}[cr][cr]{\PFGstyle $-0.6$}%
\includegraphics[width=4in]{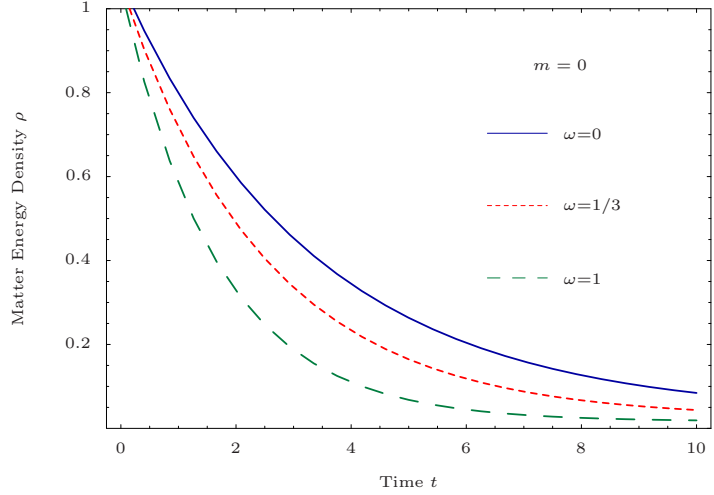}
\end{psfrags}
 \caption{Variation of matter energy density $\rho$ with cosmic time $t$ in case of $m=0$.}\protect\label{fig302}
\end{figure}
We now discuss the model (\ref{eq335}). Average scale factor $R$, expansion scalar $\theta$, Hubble parameter $H$ shear $\sigma$ and deceleration parameter $q$ are given by
\be\label{eq347}
R=e^{\beta T},
\ee
\be\label{eq348}
\theta=3H=3\beta,
\ee
\be\label{eq349}
\sigma=ke^{-3\beta T},
\ee
\be\label{eq350}
q=-1.
\ee
We obtain bulk viscous coefficient $\zeta$ and matter density $\rho$ as
\be\label{eq351}
\zeta=\zeta_0+\zeta_1\beta,
\ee
\be\label{eq352}
\rho=3\beta(\zeta_0+\zeta_1\beta)+be^{-3\beta(1+\omega)T},
\ee
where $b$ is a constant of integration. Expressions for gravitational constant $G$ and cosmological term $\Lambda$ are
\be\label{eq353}
G=-\frac{(k^2+\alpha^2e^{4\beta T})}{4\pi b(1+\omega)e^{3(1-\omega)\beta T}},
\ee
\be\label{eq354}
\Lambda=3\beta^2+\left(\frac{1-\omega}{1+\omega}\right)\frac{k^2}{e^{6\beta T}}-\left(\frac{1+3\omega}{1+\omega}\right)\frac{\alpha^2}{e^{2\beta T}}+\frac{6\beta(\zeta_0+\zeta_1\beta)(k^2+\alpha^2e^{4\beta T})}{ b(1+\omega)^2e^{3(1-\omega)\beta T}}.
\ee
The model has no initial singularity. The expansion scalar $\theta$ is constant throughout the evolution of the universe and therefore the model represents uniform expansion. Also $q=-1$ shows that the expansion of the model is accelerating with constant speed rate. At $T=0$, $R$, $\sigma$, $\zeta$, $\rho$, $G$ and $\Lambda$ all are constant. The energy density decreases as time increases and becomes constant at late times. Also $G$ and $\Lambda$ become infinite for very large values of $T$ whereas they tend to zero for $\omega<-\frac{1}{3}$ at late times. We observe that gravitational constant $G$ to be positive for $b<0$ whereas $G$ is negative for $b>0$. The possibility of negative gravitational constant $G$ has been discussed by~\citet{rf354} concluding that effective gravitational constant may have changed sign in the early universe. As $T\rightarrow\infty$, the ratio $\sigma/\theta$ becomes zero. Therefore the model approaches isotropy for large values of $T$.

\begin{figure}
\def\PFGstyle{\tiny}
\centering
\begin{psfrags}
\psfrag{mNotEqual0}[cc][cc]{\PFGstyle $m\neq 0$}%
\psfrag{Omega0}[cl][cl]{\PFGstyle $\omega =0$}%
\psfrag{Omega13}[cl][cl]{\PFGstyle $\omega =1/3$}%
\psfrag{Omega1}[cl][cl]{\PFGstyle $\omega =1$}%
\psfrag{S0}[tc][tc]{\PFGstyle $0$}%
\psfrag{S11}[tc][tc]{\PFGstyle $1$}%
\psfrag{S2}[tc][tc]{\PFGstyle $0.2$}%
\psfrag{S4}[tc][tc]{\PFGstyle $0.4$}%
\psfrag{S6}[tc][tc]{\PFGstyle $0.6$}%
\psfrag{S8}[tc][tc]{\PFGstyle $0.8$}%
\psfrag{Timet}[tc][tc]{\PFGstyle Time $t$}%
\psfrag{VacuumDens}[bc][bc]{\PFGstyle Vacuum Energy Density $\Lambda $}%
\psfrag{W0}[cr][cr]{\PFGstyle $0$}%
\psfrag{W13}[cr][cr]{\PFGstyle $100$}%
\psfrag{W22}[cr][cr]{\PFGstyle $20$}%
\psfrag{W42}[cr][cr]{\PFGstyle $40$}%
\psfrag{W62}[cr][cr]{\PFGstyle $60$}%
\psfrag{W82}[cr][cr]{\PFGstyle $80$}%
\psfrag{x0}[tc][tc]{\PFGstyle $0$}%
\psfrag{x11}[tc][tc]{\PFGstyle $1$}%
\psfrag{x151}[tc][tc]{\PFGstyle $1.5$}%
\psfrag{x21}[tc][tc]{\PFGstyle $2$}%
\psfrag{x251}[tc][tc]{\PFGstyle $2.5$}%
\psfrag{x2}[tc][tc]{\PFGstyle $0.2$}%
\psfrag{x4}[tc][tc]{\PFGstyle $0.4$}%
\psfrag{x5}[tc][tc]{\PFGstyle $0.5$}%
\psfrag{x6}[tc][tc]{\PFGstyle $0.6$}%
\psfrag{x8}[tc][tc]{\PFGstyle $0.8$}%
\psfrag{xm11}[tc][tc]{\PFGstyle $-1$}%
\psfrag{xm5}[tc][tc]{\PFGstyle $-0.5$}%
\psfrag{y0}[cr][cr]{\PFGstyle $0$}%
\psfrag{y11}[cr][cr]{\PFGstyle $1$}%
\psfrag{y21}[cr][cr]{\PFGstyle $2$}%
\psfrag{y2}[cr][cr]{\PFGstyle $0.2$}%
\psfrag{y31}[cr][cr]{\PFGstyle $3$}%
\psfrag{y41}[cr][cr]{\PFGstyle $4$}%
\psfrag{y4}[cr][cr]{\PFGstyle $0.4$}%
\psfrag{y5}[cr][cr]{\PFGstyle $0.5$}%
\psfrag{y6}[cr][cr]{\PFGstyle $0.6$}%
\psfrag{ym11}[cr][cr]{\PFGstyle $-1$}%
\psfrag{ym2}[cr][cr]{\PFGstyle $-0.2$}%
\psfrag{ym4}[cr][cr]{\PFGstyle $-0.4$}%
\psfrag{ym5}[cr][cr]{\PFGstyle $-0.5$}%
\psfrag{ym6}[cr][cr]{\PFGstyle $-0.6$}%
\includegraphics[width=4in]{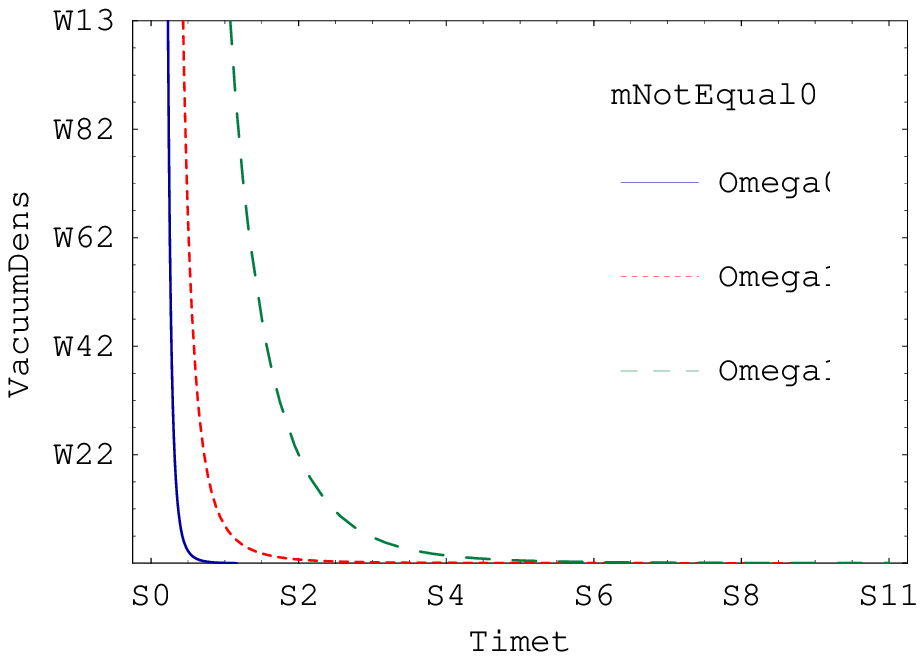}
\end{psfrags}
\caption{Variation of vacuum energy density $\Lambda$ with cosmic time $t$ in case of $m\neq0$}\protect\label{fig303}
\end{figure}
\begin{figure}
\def\PFGstyle{\tiny}
\centering
\begin{psfrags}
\psfrag{m0}[cc][cc]{\PFGstyle $m=0$}%
\psfrag{Omega0}[cl][cl]{\PFGstyle $\omega =0$}%
\psfrag{Omega13}[cl][cl]{\PFGstyle $\omega =1/3$}%
\psfrag{Omega1}[cl][cl]{\PFGstyle $\omega =1$}%
\psfrag{Omegam12}[cl][cl]{\PFGstyle $\omega =-1/2$}%
\psfrag{S0}[tc][tc]{\PFGstyle $0$}%
\psfrag{S12}[tc][tc]{\PFGstyle $10$}%
\psfrag{S22}[tc][tc]{\PFGstyle $20$}%
\psfrag{S32}[tc][tc]{\PFGstyle $30$}%
\psfrag{S42}[tc][tc]{\PFGstyle $40$}%
\psfrag{S52}[tc][tc]{\PFGstyle $50$}%
\psfrag{Timet}[tc][tc]{\PFGstyle Time $t$}%
\psfrag{VacuumDens}[bc][bc]{\PFGstyle Vacuum Energy Density $\Lambda $}%
\psfrag{W0}[cr][cr]{\PFGstyle $0$}%
\psfrag{W1}[cr][cr]{\PFGstyle $0.1$}%
\psfrag{W2m1}[cr][cr]{\PFGstyle $0.02$}%
\psfrag{W4m1}[cr][cr]{\PFGstyle $0.04$}%
\psfrag{W6m1}[cr][cr]{\PFGstyle $0.06$}%
\psfrag{W8m1}[cr][cr]{\PFGstyle $0.08$}%
\psfrag{x0}[tc][tc]{\PFGstyle $0$}%
\psfrag{x11}[tc][tc]{\PFGstyle $1$}%
\psfrag{x151}[tc][tc]{\PFGstyle $1.5$}%
\psfrag{x21}[tc][tc]{\PFGstyle $2$}%
\psfrag{x251}[tc][tc]{\PFGstyle $2.5$}%
\psfrag{x2}[tc][tc]{\PFGstyle $0.2$}%
\psfrag{x4}[tc][tc]{\PFGstyle $0.4$}%
\psfrag{x5}[tc][tc]{\PFGstyle $0.5$}%
\psfrag{x6}[tc][tc]{\PFGstyle $0.6$}%
\psfrag{x8}[tc][tc]{\PFGstyle $0.8$}%
\psfrag{xm11}[tc][tc]{\PFGstyle $-1$}%
\psfrag{xm5}[tc][tc]{\PFGstyle $-0.5$}%
\psfrag{y0}[cr][cr]{\PFGstyle $0$}%
\psfrag{y11}[cr][cr]{\PFGstyle $1$}%
\psfrag{y21}[cr][cr]{\PFGstyle $2$}%
\psfrag{y31}[cr][cr]{\PFGstyle $3$}%
\psfrag{y41}[cr][cr]{\PFGstyle $4$}%
\psfrag{y51}[cr][cr]{\PFGstyle $5$}%
\psfrag{y5}[cr][cr]{\PFGstyle $0.5$}%
\psfrag{ym11}[cr][cr]{\PFGstyle $-1$}%
\psfrag{ym5}[cr][cr]{\PFGstyle $-0.5$}%
\includegraphics[width=4in]{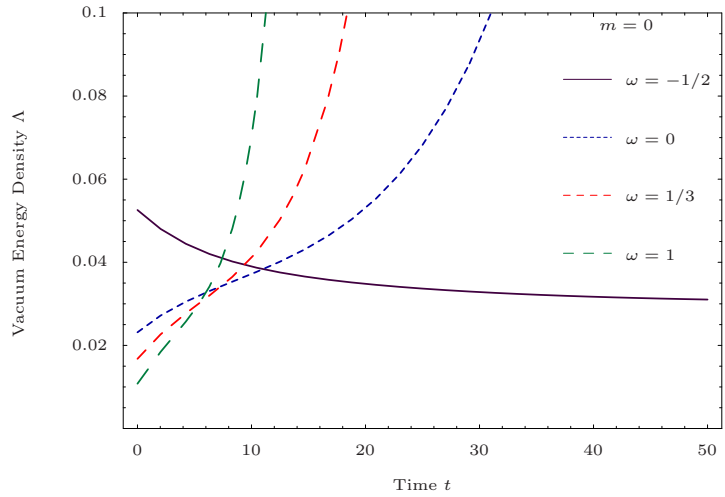}
\end{psfrags}
 \caption{Variation of vacuum energy density $\Lambda$ with cosmic time $t$ in case of $m=0$}\protect\label{fig304}
\end{figure}

\section{Conclusion}
In this paper, we have investigated spatially homogeneous and anisotropic Bianchi type V space-time with bulk viscous matter and time-dependent gravitational constant $G$ and cosmological term $\Lambda$ in general relativity. The field equations have been solved exactly by using a law of variation for the generalized Hubble parameter $H$. Two universe models have been obtained and physical behaviour of the models is discussed. Coefficient of bulk viscosity $\zeta$ is assumed to be a simple linear function of Hubble parameter $H$ (i.e. $\zeta=\zeta_0+\zeta_1 H$, where $\zeta_0$ and $\zeta_1$ are constants). For $\zeta_0=0=\zeta_1$ we recover perfect fluid models. In case of cosmology for $m\neq0$, the universe starts from a singular state whereas cosmology for $m=0$ follows a non-singular start. We observe that the presence of bulk viscosity increases the value of matter density. For the models obtained $\sigma/\theta\rightarrow0$ as $T\rightarrow\infty$. Thus the models approach isotropy at late times. From (\ref{eq331}) one concludes that for $m>1$, the model represents a decelerating universe whereas for $0\leq m<1$, it gives rise to an accelerating universe. When $m=1$, we obtain $H=\frac{1}{T}$ and $q=0$, so that every galaxy moves with constant speed. Behaviour of the matter energy density $\rho$ and vacuum energy density $\Lambda$ with respect to cosmic time are plotted in figures~\ref{fig301},~\ref{fig302},~\ref{fig303},~\ref{fig304}.



\begin{thebibliography}{}
\bibitem[\protect\citeauthoryear{Abdel-Rehman}{1990}]{rf309}
Abdel-Rehman, A.M.M. 1990, Gen. Relativ. Gravit., 22, 655

\bibitem[\protect\citeauthoryear{Abdel-Rehman}{1992}]{rf306}
Abdel-Rehman, A.M.M. 1992, Phys. Rev. D, 45, 3497

\bibitem[\protect\citeauthoryear{Abdussattar and Vishwakarma}{1997}]{rf324}
Abdussattar and Vishwakarma, R.G. 1997, Class. Quant. Grav., 14, 945

\bibitem[\protect\citeauthoryear{Arbab}{1998a}]{rf326}
Arbab, A.I. 1998a, Gen. Relativ. Gravit., 29, 391

\bibitem[\protect\citeauthoryear{Arbab}{1998b}]{rf338}
Arbab, A.I. 1998b, Gen. Relativ. Gravit., 30, 1401

\bibitem[\protect\citeauthoryear{Bali and Tinker}{2009}]{rf329}
Bali, R. and Tinker, S. 2009, Chin. Phys. Lett., 26, 029802

\bibitem[\protect\citeauthoryear{Beesham}{1986a}]{rf304}
Beesham, A. 1986a, Int. J. Theor. Phys., 25, 1295

\bibitem[\protect\citeauthoryear{Beesham}{1986b}]{rf325}
Beesham, A. 1986b, Nuovo Cimento B, 96, 17

\bibitem[\protect\citeauthoryear{Beesham et al.}{2000}]{rf339}
Beesham, A., Ghosh, S.G. and Lombard, R.G. 2000, Gen. Relativ. Gravit.,32,471

\bibitem[\protect\citeauthoryear{Berman}{1983}]{rf344}
Berman, M.S. 1983, Nuovo Cim. B, 74,182

\bibitem[\protect\citeauthoryear{Berman}{1991}]{rf303}
Berman, M.S. 1991, Gen. Relativ. Gravit., 23, 465

\bibitem[\protect\citeauthoryear{Bertolami}{1986}]{rf322}
Bertolami, O. 1986, Nuovo Cim. B, 93, 36

\bibitem[\protect\citeauthoryear{Canuto and Narlikar}{1980}]{rf307}
Canuto, V.M. and Narlikar, J.V. 1980, Astrophys. J., 6, 236

\bibitem[\protect\citeauthoryear{Carneiro}{2003}]{rf312}
Carneiro, S. 2003, Int. J. Mod. Phys. D, 12, 1669

\bibitem[\protect\citeauthoryear{Coble et al.}{1997}]{rf321}
Coble, K., Dodelson, S. and Frieman, J.A. 1997, Phys. Rev. D, 55, 1851

\bibitem[\protect\citeauthoryear{Coley}{1990}]{rf332}
Coley, A.A. 1990, Gen. Relativ. Gravit., 22, 3

\bibitem[\protect\citeauthoryear{Dirac}{1937}]{rf301}
Dirac, P.A.M. 1937, Nature, 139, 323

\bibitem[\protect\citeauthoryear{Frees et al.}{1987}]{rf319}
Frees, K. et al. 1987, Nucl. Phys. B., 287, 797

\bibitem[\protect\citeauthoryear{Frieman et al.}{1995}]{rf320}
Frieman, J.A. et al. 1995, Phys. Rev. Lett., 75, 2077

\bibitem[\protect\citeauthoryear{Garnavich et al.}{1998}]{rf315}
Garnavich, P.M. et al. 1998, Astrophys. J., 509, 74

\bibitem[\protect\citeauthoryear{Hawking and Ellis}{1975}]{rf351}
Hawking, S.W. and Ellis, G.F.R. 1975, The Large Scale Structure of Space-time, Cambridge University Press,  p. 88

\bibitem[\protect\citeauthoryear{Heller and Klimek}{1975}]{rf335}
Heller, M. and Klimek, Z. 1975, Astrophys. Space Sci., 33, 37

\bibitem[\protect\citeauthoryear{Kalligas et al.}{1992}]{rf323}
Kalligas, O., Wesson, P.S. and Everitt, C.W.F. 1992, Gen. Relativ. Gravit., 24, 351

\bibitem[\protect\citeauthoryear{Klimek}{1976}]{rf331}
Klimek, Z. 1976, Nuovo Cim. B, 35, 249

\bibitem[\protect\citeauthoryear{Knop}{2003}]{rf342}
Knop, R.A., et al. 2003, Astrophys. J., 598, 102

\bibitem[\protect\citeauthoryear{Krauss}{1995}]{rf313}
Krauss, L.M. et al. 1995, Gen. Relativ. Gravit., 27, 1137

\bibitem[\protect\citeauthoryear{Landsberg and Evans}{1977}]{rf349}
Landsberg, P.T. and Evans, D.A. 1977, Mathematical Cosmology,  Clarendon, Oxford, p. 101

\bibitem[\protect\citeauthoryear{Lau}{1985}]{rf305}
Lau, Y.K. 1985, Aust. J. Phys., 38, 547

\bibitem[\protect\citeauthoryear{Levit}{1980}]{rf308}
Levit, L.S. 1980, Lett. Nuovo Cimento, 29, 23

\bibitem[\protect\citeauthoryear{Maharaj and Naidoo}{1993}]{rf345}
Maharaj, S.D. and Naidoo, R. 1993, Astrophys. Space Sci., 208, 261

\bibitem[\protect\citeauthoryear{Meng et al.}{2007}]{rf350}
Meng, X.H., Ren, J. and Hu, M.G. 2007, Commun. Theor. Phys., 47, 379

\bibitem[\protect\citeauthoryear{Misner}{1967}]{rf333}
Misner, C.W. 1967, Nature, 214, 40

\bibitem[\protect\citeauthoryear{Murphy}{1973}]{rf334}
Murphy, J.L. 1973, Phys. Rev. D, 8, 4231

\bibitem[\protect\citeauthoryear{Ozer and Taha}{1987}]{rf318}
Ozer, M. and Taha, M.O. 1987, Nucl. Phys. B., 287, 776

\bibitem[\protect\citeauthoryear{Padmanabhan and Chitre}{1987}]{rf336}
Padmanabhan, T. and Chitre, S.M. 1987, Phys. Lett. A, 120, 433

\bibitem[\protect\citeauthoryear{Perlmutter et al.}{1999}]{rf314}
Perlmutter, S. et al. 1999, Astrophys. J., 517, 565

\bibitem[\protect\citeauthoryear{Pradhan et al.}{2004}]{rfex4}
Pradhan, A., Yadav, L. and Yadav, A.K., 2004, Czech. J. Phys., 54, 487

\bibitem[\protect\citeauthoryear{Pradhan et al.}{2005}]{rfex2}
Pradhan, A., Yadav, A.K. and Yadav, L., 2005, Czech. J. Phys., 55, 503

\bibitem[\protect\citeauthoryear{Reddy et al.}{2007}]{rf347}
Reddy, D.R.K., Naidoo, R.L. and Adhav, K.S. 2007, Astrophys. Space Sci. 307, 365

\bibitem[\protect\citeauthoryear{Riess et al.}{1998}]{rf341}
Riess, A.G. et al. 1998, Astron. J., 116, 1009

\bibitem[\protect\citeauthoryear{Riess et al.}{2004}]{rf316}
Riess, A.G. et al. 2004, Astrophys. J., 607, 665

\bibitem[\protect\citeauthoryear{Saha}{2005}]{rf337}
Saha, B. 2005, Mod. Phys. Lett. A, 20, 2127

\bibitem[\protect\citeauthoryear{Saha}{2005}]{rf328}
Saha, B. 2006, Astrophys. Space Sci., 302, 83

\bibitem[\protect\citeauthoryear{Sahni and Starobinsky}{2000}]{rf311}
Sahni, V. and Starobinsky, A.A. 2000, Int. J. Mod. Phys. D, 9, 373

\bibitem[\protect\citeauthoryear{Schmidt}{1998}]{rf317}
Schmidt, B.P. 1998, Astrophys. J., 507, 46

\bibitem[\protect\citeauthoryear{Singh and Beesham}{2010}]{rf327}
Singh, C.P. and Beesham, A. 2010, Int. J. Mod. Phys. A, 25, 3825

\bibitem[\protect\citeauthoryear{Singh and Kumar}{2009}]{rf346}
Singh, C.P. and Kumar, S. 2009, Int. J. Theor. Phys., 48, 2401

\bibitem[\protect\citeauthoryear{Singh and Baghel}{2009a}]{rf348a}
Singh, J.P. and Baghel, P.S. 2009a, Electronic J. Theor. Phys., 6, 85

\bibitem[\protect\citeauthoryear{Singh and Baghel}{2009b}]{rf348}
Singh, J.P. and Baghel, P.S. 2009b, Int. J. Theor. Phys., 48, 449

\bibitem[\protect\citeauthoryear{Singh and Baghel}{2010}]{rf340}
Singh, J.P. and Baghel, P.S. 2010, Int. J. Theor. Phys., 49, 2734

\bibitem[\protect\citeauthoryear{Singh et al.}{2008}]{rf330}
Singh, J.P., Pradhan, A. and Singh, A.K. 2008, Astrophys. Space Sci., 314, 83

\bibitem[\protect\citeauthoryear{Starobinsky}{1981}]{rf354}
Starobinsky, A.A. 1981, Sov. Astron. Lett., 7(1), 36

\bibitem[\protect\citeauthoryear{Torny et al.}{2003}]{rf343}
Tonry, J.L., et al. 2003, Astrophys. J., 594, 1

\bibitem[\protect\citeauthoryear{Weinberg}{1989}]{rf310}
Weinberg, S. 1989, Rev. Mod. Phys., 61, 1

\bibitem[\protect\citeauthoryear{Wesson}{1980}]{rf302}
Wesson, P.S. 1980, Gravity, Particles and Astrophysics (Reidel, Dordrecht).
\bibitem[\protect\citeauthoryear{Yadav}{2010}]{rfex1}
Yadav, A.K., 2010, Int. J. Theor. Phys., 49, 1140

\bibitem[\protect\citeauthoryear{Yadav et al.}{2012}]{rfex3}
Yadav, A.K., Pradhan, A. and Singh, A.K. 2012, Astrophys. Space Sci., 337, 379
\end{thebibliography}
\end{document}